\documentclass{webofc}
\usepackage[varg]{txfonts}
\allowdisplaybreaks  
\woctitle{The 13th International Symposium on Origin of Matter and Evolution of Galaxies}
\begin{document}
\title{Shedding Light on the EOS-Gravity Degeneracy and Constraining the Nuclear Symmetry Energy from the Gravitational Binding Energy of Neutron Stars}
%
%

\author{Xiao-Tao He\inst{1,2}\fnsep\thanks{\email{hext@nuaa.edu.cn}} \and
        F. J. Fattoyev\inst{2,3}\fnsep\thanks{\email{ffattoye@indiana.edu}} \and
        Bao-An Li\inst{2}\fnsep\thanks{\email{bao-an.li@tamuc.edu}} \and
        W. G. Newton\inst{2}\fnsep\thanks{\email{william.newton@tamuc.edu}}
}

\institute{College of Material Science and Technology, Nanjing University of Aeronautics and Astronautics, Nanjing 210016, China
\and
           Department of Physics and Astronomy, Texas A\&M University-Commerce, Commerce, TX 75429, USA
\and
          Center for Exploration of Energy and Matter, Indiana University, Bloomington, IN 47408, USA
          }

\abstract{%
A thorough understanding of properties of neutron stars requires both a
reliable knowledge of the equation of state (EOS) of super-dense
nuclear matter and the strong-field gravity theories simultaneously.
To provide information that may help break this EOS-gravity
degeneracy, we investigate effects of  nuclear symmetry energy on
the gravitational binding energy of neutron stars within GR and the
scalar-tensor subset of alternative gravity models. We focus on
effects of the slope $L$ of nuclear symmetry energy at saturation
density and the high-density behavior of nuclear symmetry energy. We
find that the variation of either the density slope $L$ or the
high-density behavior of nuclear symmetry energy leads to large
changes in the binding energy of neutron stars. The difference in
predictions using the GR and the scalar-tensor theory appears only
for massive neutron stars, and even then is significantly smaller
than the difference resulting from variations in the symmetry
energy. }
\maketitle

\section{Introduction}
\label{intro} 

Neutron stars (NSs) are among the most compact, dense
and neutron-rich objects known in the Universe. The typical surface
compactness parameter of NSs is of the order of $\eta_{R} \equiv
2GM/Rc^2 \approx 0.4$, which is 5 orders of magnitudes larger than
that at the surface of the sun ($\eta_{\odot} \equiv
2GM_{\odot}/R_{\odot}c^2 \approx 4\times 10^{-6}$). If one uses
instead a more natural measure of the gravitational field strength,
the ``curvature'' parameter $\xi \equiv 2GM/c^2r^3$, which is
related to the non-vanishing components of the Riemann tensor in
vacuum ${\mathcal{R}^1}_{010} = - \xi$ as well as to the Kretschmann
invariant $\mathcal{K} = 2\sqrt{3} \xi$ outside the star, the
surface curvature
for a $1.4M_{\odot}$ neutron star is then 14 orders of magnitudes
larger than
that at the surface of the Sun, where Solar System tests of General
Relativity in the weak field regime are usually performed. The
strong gravitational field makes NSs as good testbeds for the
untested strong-field gravity regime, while their high-density
matter content make them as natural laboratories for determining the
EOS of the super-dense nuclear
matter~\cite{Lattimer:2006xb,Lattimer:2012nd,Li:2008gp,Pani:2011xm,Eksi:2014wia}.
However, it cannot simultaneously be good testing ground for both.
There is a possible degeneracy between the models of the
neutron-star matter EOSs and models of gravity applied to describe
their properties. How to break this degeneracy is a longstanding
problem to which many recent studies have been devoted (see e.g.,
~\cite{Will:2005va,Harada:1998ge,Sotani:2004rq,Lasky:2008fs,Wen:2009av,
Cooney:2009rr,Horbatsch:2010hj,Arapoglu:2010rz,Deliduman:2011nw,
Pani:2011xm,Sotani:2012aa,Lin:2013rea,Yagi:2013mbt,Sotani:2014goa,
Eksi:2014wia}). This talk is based on the work originally 
published in Ref.~\cite{He:2014yqa}. Here we present our analyses of the extent to which
a EOS-Gravity degeneracy exists when models of gravity are
limited to classical GR and Scalar-Tensor (ST) theories, while
variations of the EOS appear as a result of variation of the
slope $L$ of the nuclear symmetry energy at saturation density or
the high-density behavior of the nuclear symmetry energy. We
find that the variation of either the density slope $L$ or the
high-density behavior of nuclear symmetry energy within their
uncertainty ranges lead to significant changes in the binding energy
of NSs. In particular, the variations are significantly greater than
those that result from ST theories of gravity, leading to the
conclusion that within those subset of gravity models, measurements
of neutron star properties constrain mainly the EOS. Further
investigations demonstrate that only EOSs with the soft symmetry
energy at high-density are consistent with constraints on the
gravitational binding energy of PSR J0737-3039B.

\section{The Equation of State of Nuclear Matter}
\label{sec:EOS} 

In general, the EOS of neutron-rich nucleonic matter
can be expressed in terms of the binding energy per nucleon in
asymmetric nuclear matter of isospin asymmetry
$\alpha=(\rho_n-\rho_p)/\rho$ as
$E(\rho,\alpha)=E_{0}(\rho)+\mathcal{S}(\rho)\alpha^{2}+\mathcal{O}(\alpha^{4})$,
where $E_0(\rho)$ is the binding energy per nucleon in symmetric
nuclear matter ($\alpha=0$), $\mathcal{S}(\rho)$ is the nuclear
symmetry energy, and $\rho=\rho_{n}+\rho_{p}$ is the total baryon
density with $\rho_{n}$ ($\rho_{p}$) being the neutron (proton)
density. The EOS can further be characterized in terms of bulk
nuclear properties by expanding both $E_{0}(\rho)$ and
$\mathcal{S}(\rho)$ in Taylor series around nuclear saturation
density $\rho_0 = 0.16$ fm$^{-3}$, i.e., $E_{0}(\rho)=B_0
+\frac{1}{2} K_{0} \chi^{2}+\mathcal{O}(\chi^{3})$ and
$\mathcal{S}(\rho)=J+L\chi+\frac{1}{2} K_{\rm
sym}\chi^{2}+\mathcal{O}(\chi^{3})$, where $\chi \equiv
(\rho-\rho_{0})/3\rho_{0}$ quantifies the deviations of the nuclear
density from its saturation value. Terrestrial experimental nuclear
data such as the ground state properties of finite nuclei and
energies of giant resonances tightly constrain the binding energy at
saturation $B_0$ and the nuclear incompressibility coefficient
$K_{0}$, hence constrain the EOS of symmetric nuclear matter
$E_{0}(\rho)$. Whereas the symmetry energy at saturation $J$ is more
or less known, its density slope $L$ is largely unconstrained, and
the overall behavior of the symmetry energy at supra-saturation
densities is quite uncertain.
\begin{figure}[h]
\sidecaption
\centering
\includegraphics[width=2.2in,angle=0]{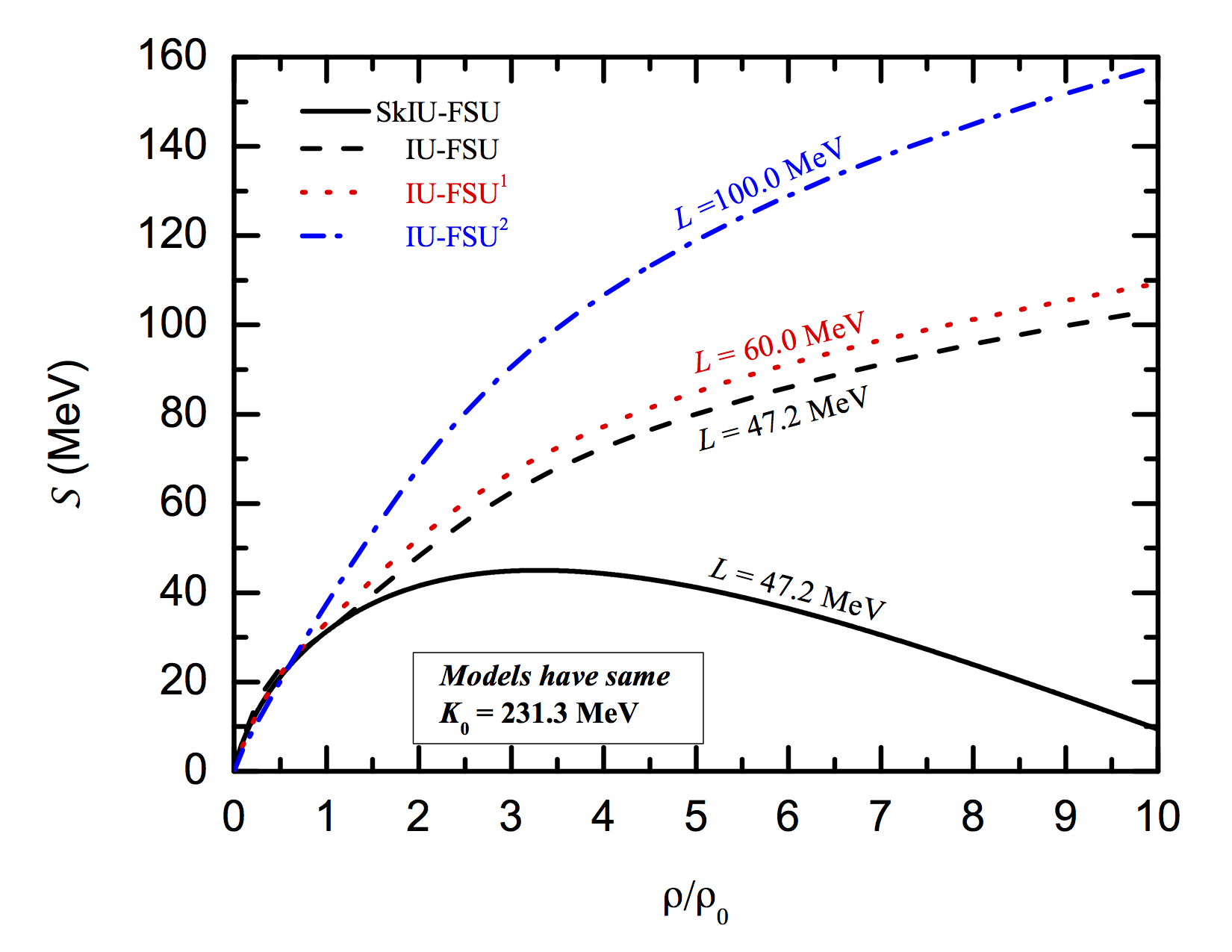}
\caption{(color online). Density dependence of the nuclear symmetry
energy: the solid black line corresponds to the symmetry energy
predictions in SkIU-FSU, the dashed black line corresponds to the
original IU-FSU model, the dotted red and dash-dotted blue lines are
symmetry energy predictions in the IU-FSU-like models with density
slope $L=60.0$ MeV (IU-FSU$^1$) and $L=100.0$ MeV (IU-FSU$^2$),
respectively. Figure is taken from Ref.~\cite{He:2014yqa}.}.
\label{Fig:EOS}
\end{figure}

For our base models we use two recently established EOSs for
neutron-rich nucleonic matter within the IU-FSU Relativistic Mean
Field (RMF) and the SkIU-FSU Skyrme-Hartree-Fork (SHF)
models~\cite{Fattoyev:2010mx,Fattoyev:2012ch,Fattoyev:2012uu}. The
slope of the symmetry energy at saturation for these models is $L =
47.2$ MeV. While the density dependence of symmetry energy
$\mathcal{S}(\rho)$ at subsaturation densities is almost identical
for these models, at supra-saturation densities of $\gtrsim
1.5\rho_0$ the two models give significantly different predictions
of $\mathcal{S}(\rho)$. To test the sensitivity of the gravitational
binding energy of neutron stars to the variations of properties of
neutron-rich nuclear matter around saturation density, we have
further introduced two RMF models with density slopes of the
symmetry energy at saturation density of $L=60$ MeV and $L=100$ MeV.
Shown in Fig.~\ref{Fig:EOS} are the density dependence of the
symmetry energy for these four models~\cite{He:2014yqa}.

\section{Comparison of Predictions between GR \& Scalar-Tensor Theory Using Different EOSs}
\label{sec:Binding Energy}

Properties of neutron stars are not only sensitive to the underlying
EOS, but also to the strong-field behavior of gravity. We will
consider neutron stars in both GR and in the simplest natural
extension of the GR known as the {\sl scalar-tensor theory} of
gravity. According to this theory, in addition to the second-rank
metric tensor, $g_{\mu\nu}$, the gravitational force is also
mediated by a scalar field, $\varphi$. The action defining this
theory can be written in the most general form
as~\cite{Damour:1996ke,Yazadjiev:2014cza}
\begin{equation}
S = \frac{c^4}{16 \pi G} \int d^4x \sqrt{-g^{\ast}}\left[R^{\ast} -2
g^{\ast \mu \nu} \partial_{\mu} \varphi \partial_{\nu} \varphi -
V(\varphi)\right] + S_{\rm matter}\left(\psi_{\rm matter};
A^2(\varphi)g^{\ast}_{\mu\nu} \right) \ ,
\end{equation}
where $R^{\ast}$ is the Ricci scalar curvature with respect to the
so-called {\sl Einstein frame} metric $g^{\ast}_{\mu\nu}$ and
$V(\varphi)$ is the scalar field potential. The physical
metric is defined as
$g_{\mu\nu} \equiv A^2(\varphi)g^{\ast}_{\mu\nu}$. The relativistic
equations for stellar structure in hydrostatic equilibrium can be
written as
\begin{eqnarray}
\frac{dM_{\rm G}(r)}{dr} &=& \frac{4 \pi}{c^2} r^2 \mathcal{E}(r)
A^4(\varphi) +
\frac{r^2}{2}\left(1-\frac{2GM(r)}{c^2r}\right)\chi^2(r) +
\frac{r^2}{4} V(\varphi) \ , \\
\frac{dM_{\rm B}(r)}{dr} &=& 4\pi m_{\rm B} r^2 \rho(r) A^3(\varphi) \ , \\
\frac{d \varphi(r)}{dr} &=& \chi(r) \ , \\
\nonumber \frac{d \chi(r)}{dr} &=& \Bigg[1-\frac{2 G M(r)}{c^2
r}\Bigg]^{-1}\Bigg\{\frac{4 \pi G}{c^4} A^4(\varphi)
\bigg[\alpha(\varphi) \bigg(\mathcal{E}(r) - 3 P(r)\bigg)+r\chi(r)\bigg(\mathcal{E}(r) - P(r)\bigg)\bigg] \ \\
&-& \frac{2}{r}\left(1-\frac{ G M(r)}{c^2 r}\right) \chi(r)
+\frac{1}{2}r\chi(r)V(\varphi) +
\frac{1}{4}\frac{dV(\varphi)}{d\varphi}\Bigg\} \ , \\
\nonumber \frac{dP(r)}{dr} &=& -\bigg(\mathcal{E}(r) +
P(r)\bigg)\Bigg[1-\frac{2 G M(r)}{c^2 r}\Bigg]^{-1}\Bigg\{ \frac{4
\pi G}{c^4} r A^4(\varphi) P(r)  + \frac{G}{c^2r^2}M(r)  \ \\
&+& \left(1-\frac{2 G M(r)}{c^2 r}\right)\left(\frac{1}{2}r\chi^2(r)
+ \alpha(\varphi)\chi(r)\right) -\frac{1}{4}rV(\varphi) \Bigg\} \ ,
\end{eqnarray}
where $\alpha(\varphi) \equiv \partial \ln A(\varphi)/\partial
\varphi$. Following the Ref.~\cite{Damour:1996ke} we set
$V(\varphi)=0$ which can only appear in models of modified gravity,
and we consider a coupling function of the form $A(\varphi) =
\exp\left(\alpha_0\varphi + \frac{1}{2}\beta_0
\varphi^2\right)$~\cite{He:2014yqa}. For a given central pressure
$P(0) = P_{\rm c}$ one can integrate the equations above from the
center of the star to $r \rightarrow \infty$, where the only input
required is the EOS of dense matter in chemical equilibrium. At the
center of the star the Einstein frame boundary conditions are given
as $P(0) = P_{\rm c} \ , \mathcal{E}(0) = \mathcal{E}(\rm c) \
,\varphi(0) = \varphi_{\rm c} \ , \chi(0) = 0$, while at infinity we
demand cosmologically flat solution to agree with the observation
$\lim_{r \rightarrow \infty} \varphi(r) = 0$. The stellar coordinate
radius is determined by the condition of $P(r_{\rm s}) = 0$. The
physical radius of a neutron star is found in the Jordan frame as
$R_{\rm NS} = A^2\left[\varphi(r_{\rm s})\right]r_{\rm s}$. Notice
however that the physical stellar mass as measured by an observer at
infinity matches with the coordinate mass, since at infinity the
coupling function approaches unity. According to the latest
observational constraints the quadratic parameter of the
scalar-tensor theory should take values of not smaller than $\beta_0
\gtrsim -5.0$~Ref.~\cite{Freire:2012mg,Damour:1996ke}. Similarly the
absolute value of the linear parameter is also constrained very well
by observation. We will therefore choose the upper bounds on these
parameters as $\alpha_0^2 < 2.0 \times 10^{-5}$ and $\beta_0 >
-5.0$~\cite{Freire:2012mg}.

\begin{figure}[h]
\sidecaption
\includegraphics[width=2.5in,angle=0]{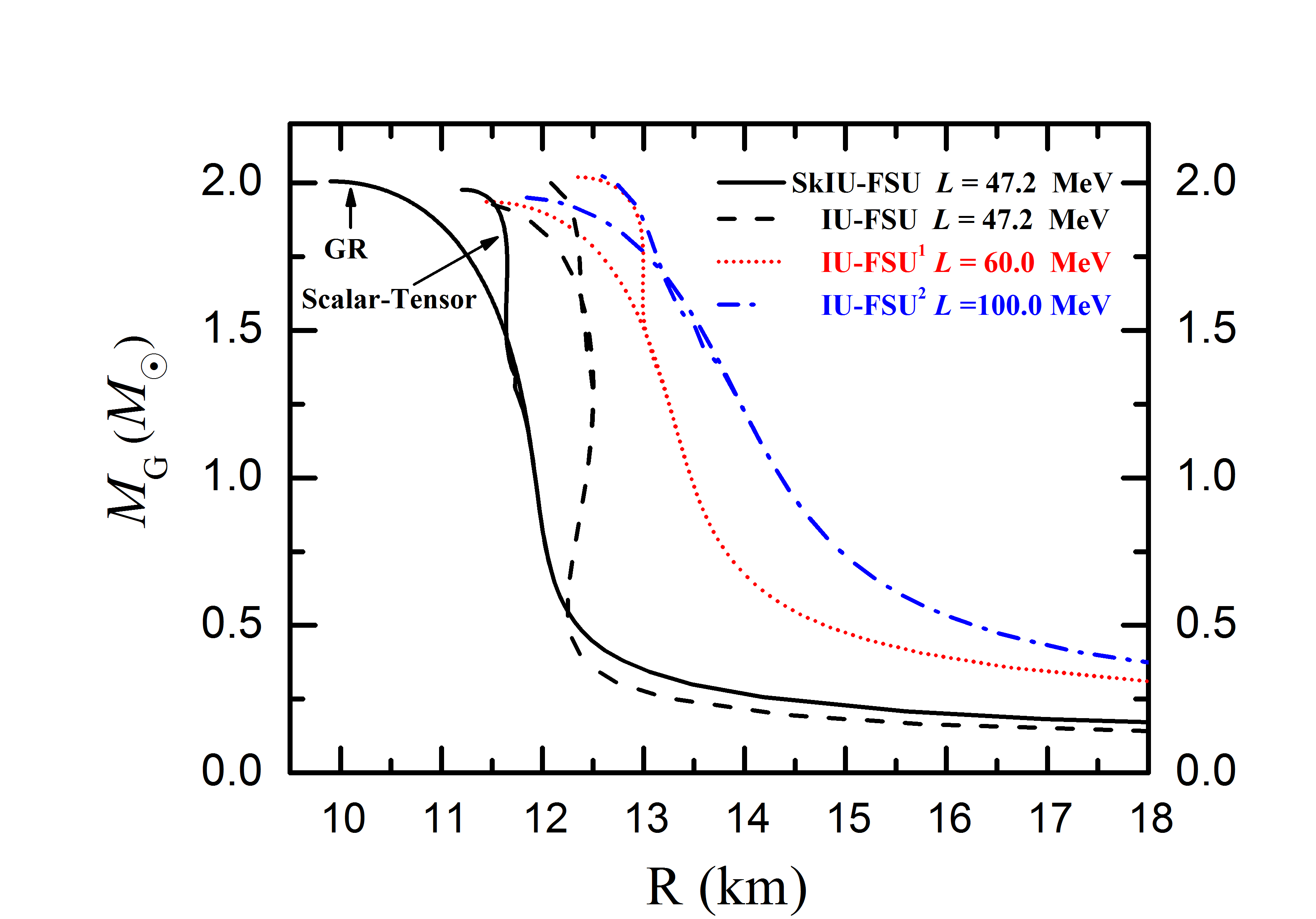}
\caption{(color online). The mass-versus-radius relation of neutron
stars calculated using EOSs considered in this work. For the
scalar-tensor theory  the upper observational bound on parameters
$\{\alpha_0,\beta_0\} = \{\sqrt{2.0 \times 10^{-5}}, -5.0\}$ have
been used. Figure is taken from Ref.~\cite{He:2014yqa}.}
\label{Fig:MR}
\end{figure}

In Fig.~\ref{Fig:MR} we present our results for the mass versus
radius relation of neutron stars as calculated by using the four
EOSs discussed above both within GR and the scalar-tensor theory
with $\{\alpha_0,\beta_0\} = \{\sqrt{2.0 \times 10^{-5}}, -5.0\}$. we observe a larger
radius in the scalar-tensor theories than in GR only for massive
neutron stars. Even so, the changes in radii are much smaller than
those that arise from variation of the stiffness of the symmetry
energy. In general, differences in predictions using the GR and the
scalar-tensor theories within current observational bound on their
parameters are much smaller than those due to the uncertainties in
the EOS.


\section{Constraining the Nuclear Symmetry Energy from the Gravitational Binding Energy}
\begin{figure}[h]
\sidecaption
\includegraphics[width=2.5in,angle=0]{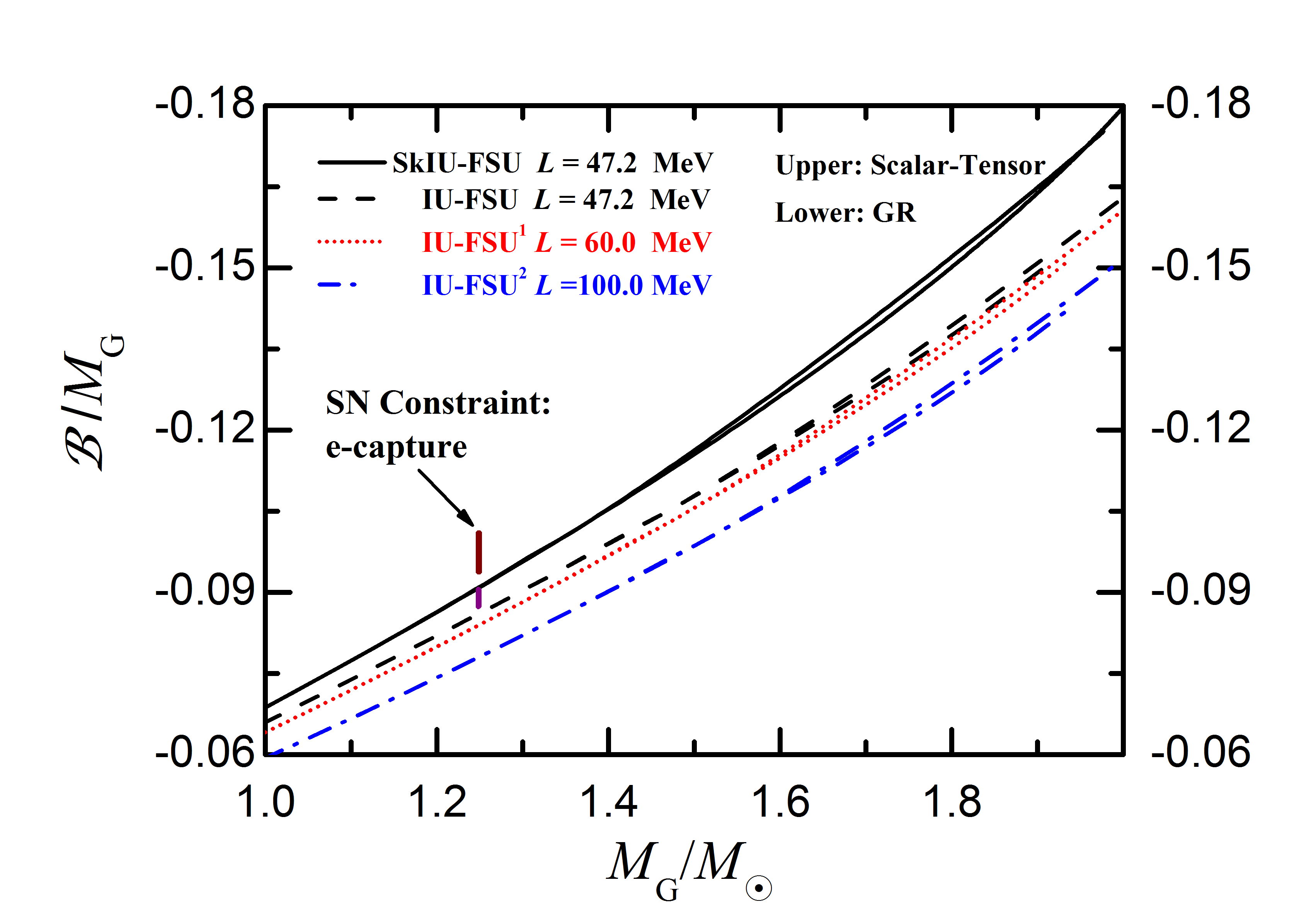}
\caption{(color online). The fractional gravitational binding energy
versus the gravitational mass of a neutron star for a set of the
EOSs discussed in the text. The constraints on the gravitational
binding energy of PSR J0737-3039B, under the assumption that it is
formed in an electron-capture supernova, are given by the vertical
brown and purple lines. Figure is taken from
Ref.~\cite{He:2014yqa} with slight modifications.} \label{Fig:BE}
\end{figure}

The fractional gravitational binding energy $\mathcal{B}\equiv
M_{\rm G}-M_{\rm B}$~\cite{Weinberg:1972} [See Eqns.(2)-(3)] as a
function of the total gravitational mass of a neutron star is
displayed in Fig.~\ref{Fig:BE}. Notice that in general the GR
predictions give a lower absolute value of the fractional binding
energy for a given NS mass then the scalar-tensor theory. However
the difference is quite negligible compared to the uncertainties
coming from the variations of the EOS. Furthermore, all low-mass NSs
are indistinguishable for an observer in these two models of
gravity, because the critical value for the so-called ``spontaneous
scalarizaton'' is reached only when NSs masses exceed about
$1.4$ solar mass. Thus measurements of NS mass and radii will lead
to significantly constrain density dependence of the symmetry
energy, rather than constraining gravity models within the GR and
scalar-tensor theories.

As both the mass-verus-radius relation and
gravitational binding energy of NSs are more sensitive to different
EOSs, we further examine the effects on binding energy of the
uncertain symmetry energy in more details (see again
Fig.~\ref{Fig:BE}). Notice that with the very similar low-density
symmetry energy up to about $1.5\rho_0$, but different high-density
behaviors, predictions for the $\mathcal{B}/M_{\rm G}$ in the
SkIU-FSU and IU-FSU models are quite different. This effect becomes
more evident with the increase of the total gravitational mass. With
the IU-FSU  as a reference baseline model then the relative changes
in the fractional gravitational binding energies are $5.32\%$,
$6.52\%$ and $9.89\%$ for $1.25M_{\odot}$, $1.4M_{\odot}$ and
$1.9M_{\odot}$ neutron stars respectively. This is in contrast to
the predictions of the binding energy in models with soft and stiff
symmetry energies \emph{at saturation}---{\sl e.g.}, the IU-FSU with
$L=47.2$ MeV and the IU-FSU$^2$ with $L=100$ MeV. Again using the
former as a reference model, then the relative changes in the
fractional gravitational binding energies are $-10.26\%$, $-8.85\%$
and $-7.44\%$ for $1.25M_{\odot}$, $1.4M_{\odot}$ and $1.9M_{\odot}$
neutron stars respectively. Hence the fractional binding energy is
more sensitive to the saturation density slope of the symmetry
energy for low-mass neutron stars~\cite{Newton:2009vz}, while more
sensitive to the supra-saturation behavior of the symmetry energy
for massive stars~\cite{He:2014yqa}.

The gravitational mass of the lighter pulsar PSR J0737-3039B is
determined very accurately to be $M_{\rm G} = 1.2489 \pm
0.0007M_{\odot}$~\cite{Kramer:2006nb}, which is one of the lowest
reliably measured mass for any neutron star up to date. Due to its
low mass it was suggested that this NS might have been formed as a
result of the collapse of an electron-capture supernova from a an
O-Ne-Mg white dwarf progenitor~\cite{Podsiadlowski:2005ig}. Under
this assumption, it is estimated that the baryonic mass of the
precollapse O-Ne-Mg core should lie between $1.366 M_{\odot} <
M_{\rm B} < 1.375M_{\odot}$~\cite{Podsiadlowski:2005ig} and $1.358
M_{\odot} < M_{\rm B} < 1.362M_{\odot}$~\cite{Kitaura:2005bt}. In
Fig.~\ref{Fig:BE} we show that only models with the soft symmetry
energy are consistent with these set of constraints. 
\section{Summary}
\label{sec:Conclusion} Interpreting properties of neutron stars
require a resolution in the degeneracy between the EOS for
super-dense matter and the strong-field gravity. With the goal of
providing information that may help break this degeneracy we have
studied effects of the nuclear symmetry energy within its current
uncertain range on the mass-versus-radius relation and the binding
energy of NSs within both the GR and the scalar-tensor theory of
gravity. We have found that radii of neutron stars are primarily
sensitive to the underlying EOS through the density dependence of
the symmetry energy. Within the simplest natural extension of the GR
known as the scalar-tensor theory of gravity, and by using upper
observational bounds on the parameters of the theory, we found a
negligible change in the binding energy of NSs over the whole mass
range, and significant changes in radii only for NSs whose mass are
well above $1.4M_{\odot}$. Even then, the changes in radii are found
to be much smaller than those that result from variation of the
stiffness of the symmetry energy. We have also shown that the
gravitational binding energy is moderately sensitive to the nuclear
symmetry energy, with lower mass neutron stars $\lesssim
1.4M_{\odot}$ probing primarily the stiffness of the symmetry energy
at nuclear saturation density, and massive stars being more
sensitive to the high density behavior of the symmetry energy. A
combination of observational and theoretical arguments on the
gravitational binding energy of PSR J0737-3039B imply that only EOSs
with soft symmetry energy at high density are consistent with the
extracted values of the binding energy.

\section{Acknowledgements}
This work is supported in part by the National Natural Science
Foundation of China under Grant No. 11275098, 11275067 and
11320101004, the US National Science Foundation under Grant No.
PHY-1068022 and the CUSTIPEN (China-U.S. Theory Institute for
Physics with Exotic Nuclei) under DOE grant number
DE-FG02-13ER42025, and DOE grants DE-SC0013702 (Texas A\&M
University-Commerce), DE-FG02-87ER40365 (Indiana University) and
DE-SC0008808 (NUCLEI SciDAC Collaboration).

\bibliography{ReferencesFJF}
\vfill\eject

\end{document}